# Layer-polarized Transport via Gate-defined 1D and 0D PN Junctions in Double Bilayer Graphene

Wei Ren[1], Xi Zhang[1], Shiyu Guo[1], Jeongsoo Park[1], Jack Tavakley[1], Daochen Long[1], Kenji Watanabe[2], Takashi Taniguchi[3], Ke Wang[1*]

[1]*School of Physics and Astronomy, University of Minnesota, Minneapolis, Minnesota 55455, USA*
[2]*Research Center for Electronic and Optical Materials, National Institute for Materials Science, 1-1 Namiki, Tsukuba 305-0044, Japan*
[3]*Research Center for Materials Nanoarchitectonics, National Institute for Materials Science, 1-1 Namiki, Tsukuba 305-0044, Japan*

**We fabricate twisted double bilayer graphene devices with zero twist angle and a set of local top and bottom gates aligned perpendicularly to each other. A 1D PN junction can be electrostatically defined when the gate voltages applied to the top gates are the same but different on the bottom gates. Resistance peaks are observed at finite doping instead of at the charge neutrality points, exhibiting an unconventional broken-cross shape that arises from layer polarization of the P and N region, which can be further enhanced with finite magnetic fields. A 0D point junction (PJ) can be electrostatically defined by applying different gate voltages to the top and bottom gates, such that the P and N sides of the device are connected at a single point in the center of the device. As finite magnetic field B increases, the quantum Hall (QH) states are selectively brought into contact or away from each other depending on their layer polarization, leading to unconventional quantum oscillations which characterize the layer-polarized band-crossing. Our work provides new insights into understanding band-structure evolution and layer polarization in twisted bilayers and paves the way for new device functionality based on manipulating layer-polarized electronic states.**

The novel superconductivity and correlated insulator behaviors observed in twisted bilayer graphene (tBLG) with θ ~ 1.1° open a new direction to investigate the moiré flat band system [1–7]. Recently, similar unconventional behaviors have been found in twisted transition dichalcogenide metal (tTMD) [8–13] and the twisted double bilayer graphene (tDBLG) systems [14–17]. The band structure of tDBLG can be tuned by a perpendicular electric field (or displacement field D), and its flat band is therefore tunable without requiring precise twist angle alignment [18]. Electrostatic screening plays a stronger role in tDBLG compared to tBLG due to increased atomic thickness, as demonstrated in tDBLG with moderate (~ 2.4°) and large (~ 5°) twisted angles [19,20] and significant layer polarization.

Small-angle tBLG and tDBLG have recently attracted significant research interest due to their large moiré unit cells, which are ideal for imaging [21–23]. However, electrical transport studies of tDBLG with small or zero interlayer twist angle remain relatively unexplored. In this work, we study tDBLG with zero twist angle, hereafter referred to as B0B. In this material platform, the bands of the two participating BLG pieces have zero offset in k-space, but are offset in energy due to strong interlayer screening tunable by the displacement field. We characterize the resulting layer polarization using a conventional 1D PN junction [24,25], electrostatically defined on B0B. The measured resistance peaks of the PN junction exhibit an unconventional broken-cross shape, and the offset from the charge neutrality point quantitatively characterizes the layer polarization due to the energy offset of the two BLG bands. Hybridization between the layer-polarized bands resulted into a reconstructed tDBLG band structure with a sombrero-like band-crossing, reminiscent of that observed in BLG [26–30]. We show that such band-crossing can be electrostatically tuned and characterized via layer-polarized QH tunneling across an electrostatically defined 0D PJ [31]. Our device scheme provides new insights into characterizing the interlayer screening, layer polarization and sombrero-like band crossing.

The schematic of the device architecture is shown in Fig. 1(a). Two pieces of BLG are stacked with zero twist angle, encapsulated by atomically clean hexagonal boron nitride (hBN) flakes. The B0B stack is subsequently transferred onto a pair of 1-μm-wide bottom metal gates aligned in parallel and separated by ~100 nm ($B_1$ and $B_2$). A similar pair of 1-μm-wide metal top gates ($T_1$ and $T_2$) are deposited on the stack but oriented orthogonally to the bottom gates. The dual-gated structure allows the charge carrier density and perpendicular displacement field to be independently tuned in four regions of the device, allowing versatile electrostatic configurations for a uniform 2DEG, 1D PN junction or 0D PJ, which we utilize to comprehensively characterize the layer-polarized electronic transport in this material and device platform.

We start by measuring a uniform B0B 2DEG, with the same voltages applied to adjacent gates ($V_t = V_{t1} = V_{t2}$, $V_b = V_{b1} = V_{b2}$). The carrier density $n$ of the entire device is uniform and given by $n = C_t V_t + C_b V_b$, and the perpendicular displacement field $D$ is given by $D/\varepsilon_0 = (C_b V_b - C_t V_t)/2$. Here, the positive $D$ field direction is defined as pointing from the bottom to the top layer. Even at $n = 0$ where B0B is overall charge-neutral, a finite D field can result in electron (hole) accumulation in bottom (top) BLG due to interlayer screening, as charge density in the bottom (top) BLG is predominantly determined by the bottom (top) gate voltages [Fig. 1(b), left]. The inter-layer screening length in graphite is approximately 7 Å, equivalent to the thickness of 2–3 graphene layers, resulting in an out-of-the-plane charge distribution on a similar length scale [Fig. 1(b), right]. With four atomic layers, the B0B material platform is thick enough for screening to play an important role, yet thin enough to retain reduced but finite capacitive coupling between a gate and its furthest atomic layers. The electric field from the top (bottom) gate can still tune carrier density in the bottom (top) BLG, though less effectively due to significant (but not complete) screening from the top (bottom) BLG.

Under a large $D$ field and with strong interlayer screening, a significant charge imbalance between two layers leads to an offset between two individual BLG bands [Fig. 1(c), left], exceeding the displacement-field-induced bandgap. The valence band of the top layer and the conduction band of the bottom layer cross, and the reconstructed band structure exhibits sombrero-like dispersion bands near the charge neutrality [Fig. 1(c) right].

Fig. 1(d) shows the measured four-probe resistance of the uniform B0B 2DEG as a function of carrier density $n$ and displacement field $D$. Local resistance maxima (green and red dots) are observed at the $n = 0$ charge neutrality point, corresponding to the Fermi level within the gap of the reconstructed B0B band structure. The system is depleted of charge carriers, and charge transport is conducted by intrinsic charge carriers, as confirmed by the temperature dependence of $n = 0$ resistance peak. The extracted $n = 0$ bandgap $\Delta$ increases with the size of the applied $D$ field [Fig. 1(e)], due to interlayer band crossing with larger energy offset [see Supplementary Material (SM) for more details]. Additional resistance peaks are observed at finite doping. For example, at $D > 0$ and $n < 0$, a side resistance peak with much lower peak resistance (yellow dot) is observed, corresponding to a layer-polarized charge configuration of hole-type charge carriers predominantly residing in the top BLG, while the bottom BLG is nearly depleted of charge carriers. This also qualitatively corresponds to [Fig. 1(d), top left inset] the Fermi level aligning with the top layered-polarized dispersion of reconstructed band structure, while residing within the gap of the dispersion arising from the bottom layer. Similar side peaks are observed at the $n > 0$ and $D < 0$, with electron-type carriers polarized in the bottom layer. However, side resistance peaks are relatively ill-defined when $n$ and $D$ are of the same sign, potentially due to a more asymmetric band reconstruction due to the complex landscape of local atomic relaxation.

A 1D lateral junction can be electrostatically defined across the boundary of the two independently gated regions [Fig. 2(a)], by applying the same voltage to the two top gates $V_t = V_{t1} = V_{t2}$, but different gate voltages $V_{b1}$ and $V_{b2}$ to the individual bottom gates, which independently tune the carrier density $n_1$ and $n_2$ in each region, respectively. Fig. 2(b) shows the measured four-probe resistance as a function of $n_1$ and $n_2$, taken at a constant top gate voltage of $V_t = V_{t1} = V_{t2} = -7.8$ V and a zero magnetic field (see SM for the dependence on $V_t$).

In previously-reported graphene or tBLG PN junctions [24,25,32,33], the local resistance peaks are observed when either of the two regions reaches charge neutrality, exhibiting a cross-shape formed by two perpendicular lines at $n_1 = 0$ and $n_1 = 0$ across the entire measurement range [Fig. 2(b), dashed], marking the boundaries between the PN, PP, NP, and NN junction configuration. In contrast, the resistance peaks of B0B PN junction exhibit a broken-cross shape. The local resistance peaks are found away from the charge neutrality point – along lines of finite $n_1$ and $n_2$ – and shift into (instead of marking the boundaries of) the domains in which the device is configured as a PN or NP junction.

This shift arises from the layer polarization in B0B. For signature high resistance points in transport data, the corresponding out-of-plane charge distribution in each device region is depicted (with the same markers) in Fig. 2(c). For example, within the PN domain (top-left quarter) of Fig. 2(b), the highest resistance is observed (triangle) when the P-type carrier of $n_1 < 0$ is polarized in the top BLG of region 1, and N-type carrier of $n_2 > 0$ is polarized in the bottom BLG of region 2 [Fig. 2(c), triangle]. Electrical conductance requires both lateral (across PN junction) and vertical (from top to bottom BLG) tunneling (curved arrows), which also comprises a device configuration with the lowest conductance. At this configuration, signature of strong interlayer Coulomb-drag has also been observed (see SM), confirming the formation of a PN junction that is both lateral and vertical, across which the carrier type and layer-polarization simultaneously switch sign. This can potentially serve as a new device platform to facilitate correlated electronic behavior such as exciton condensation [34–36], enabled by significantly enhanced strong Coulomb coupling between electronic states separated by just ~1 nm.

Starting from the layer-polarized PN junction configuration, increasing P-type (N-type) carrier density in region 1 (2) will start to slightly dope the previously depleted layer, resulting in slight reduction in layer-polarization, as depicted by the square (pentagon) in Fig. 2(c). This increases both the overlap of vertical carrier density span between P and N regions and the channel for electrical conductance via lateral tunneling alone (straight arrows), resulting in a reduction in the measured peak resistance. Starting from these two configurations and moving away from the broken-cross by increasing carrier density in the other device regions, both regions are unpolarized with significant overlap of vertical charge distribution. The lateral conductance channel is opened within graphene layers, resulting in comparably effortless transport with high conductance, similar to a conventional graphene PN junction. Similar transport features with same amount of offset from the charge neutrality lines are observed in the NP configuration of the device (circle, diamond, trapezoid), corresponding to formation of the NP junction with opposite layer polarizations, further confirming the physics picture.

At a high magnetic field $B = 6$ T [Fig. 2(d)], due to Landau quantization, the lightly-doped graphene layers become more insulated, and the resistance peaks become sharper. Fig. 2(e) shows four signature 1D cuts, plotted as a function of $\Delta n$, with $\Delta n = 0$ defined at charge neutrality point, and positive $\Delta n$ direction along the scan direction marked by while arrows. All four cuts have similar peak position at $\Delta n = (1.0 \pm 0.1) \times 10^{11}$ cm$^{-2}$. This carrier density quantitatively characterizes the charge imbalance between the top and bottom BLG layers due to the strong interlayer screening.

By applying unique gate voltages to each of the four gates, the device is electrostatically divided into four different regions [Fig. 3(a)], each with unique carrier density $n$ and displacement field $D$ determined by the top (bottom) gate above (below), respectively. At the gate voltage combination of $(V_{t1}, V_{t2}, V_{b1}, V_{b2}) = (+V_0, -2V_0, +2V_0, -V_0)$, in which $V_0$ is a positive number, the bottom left (top right) region is N (P) doped, and the top left (bottom right) region is charge neutral ($n = 0$) and in gap with a positive (negative) displacement field. At this configuration, a 0D PJ is established by bringing P and N region into contact at the center of the device (or the common corner of all four regions), where both the carrier density $n$ and displacement field $D$ reach zero and switches sign simultaneously.

To characterize the layer-polarized band-crossing, we measure the four-probe magnetoresistance across the PJ as a function of magnetic field $B$ [Fig. 3(b)] and displacement field $D$. As an example for discussing the device operation and physics picture, we plot a 1D cut (along the green dashed line, at $D/\varepsilon_0 = 0.3$ V/nm) of measured four-probe resistance $R$ versus magnetic field, taken at temperatures ranging from 1.5 K to 32 K (in 4 K steps, except from 1.5 K to 4 K).

In P and N regions near the vicinity of the PJ, the B0B is lightly doped with two co-existing Fermi surfaces: a larger Fermi surface belonging to the outer bands, and a smaller Fermi surface belonging to the inner bands of the sombrero-like band. Under a high magnetic field, such as at $B = 4$ T [Fig. 3(c), square], the two fermi surfaces give rise to two types of QH edge states in both the N and P region. Electrical transport across the device is facilitated by tunneling across inner-most QH edge states [37], with tunneling resistances increasing rapidly as temperature decreases [Fig. 3(c), square]. As the Landau gap increases with higher magnetic fields, inside N region and near the PJ, the N-type (P-type) Landau level shifts away from (towards) the band gap, resulting in corresponding N-type (P-type) QH edge states moving away (towards) the charge neutrality point at the center of the device, where PJ is located [Fig. 3(d), square]. Similar, P-type (N-type) QH edge states in the P region also moves away (towards) the PJ as the magnetic field increases.

The QH edge states that move towards the PJ (and each other) belong to the inner sombrero-like bands of the P and N regions. As the magnetic field increases, the P-type QH edge state in the N region and the N-type QH edge state in the P region approach each other and eventually come into direct contact at PJ at $B = 4.8$ T. The electrical transport across the PJ device therefore does not require tunneling [Fig. 3(d), star], exhibiting a local resistance minimum and nearly zero temperature dependence.

No further semi-metallic resistance dips are observed beyond $B = 4.8$ T. This suggests that the $B = 4.8$ T dip corresponds to the last inner-band QH edge states merging at the center of the PJ [Fig. 3(c), star]. The magnetic field at which this dip is observed [Fig. 3(b), white-dashed line] corresponds to the inner Landau levels being pushed all the way down to the bottom of the sombrero located at the band edges [Fig. 3(c), star], from which the energy of the sombrero [Fig. 3(e)], as a function of $D$ field can be characterized (see SM).

As the magnetic field continues to increase beyond $B = 4.8$ T (circle), the inner bands are depleted of Landau levels, and P-type (N-type) QH edge states are no longer present in N (P) regions [Fig. 3(d), circle]. Transport across the PJ solely relies on the tunneling of the remaining QH edge states from the outer bands, whose separation rapidly increases with $B$. As a result, the measured PJ resistance increases sharply with an increasing $B$ and a decreasing $T$ [Fig. 3(c), circle].

To further verify this picture, the device is reconfigured into $(-V_0, -2V_0, +2V_0, +V_0)$ so that a finite tunneling barrier is established at the PJ rather than a band gap closure (Fig. S2). This prevents direct contact between inner-band QH edge states, causing resistance minima to become ill-defined and strongly temperature-dependent beyond $B = 4$ T, consistent with conductance being limited by tunneling (see SM for details).

**Conclusion**

In conclusion, we study zero-twist double-bilayer graphene, in which the strong interlayer screening results in layer polarization and tunable interlayer band crossing. We fabricate nano-devices based on this material platform with two pairs of top and bottom local gates orthogonally aligned, allowing the device to be versatilely configured as a homogeneous 2DEG, a 1D PN junction, or a 0D PJ to manipulate the layer degree of freedom. We demonstrate versatile and tunable layer-polarized quantum transport via a 1D PN junction and 0D PJ, and quantitatively characterize the layer polarization and the resulting band crossing. Our work provides a new approach to engineer band structure by hybridizing two material bands with energy offset via screening instead of crystal momentum offset via twisting. The layer-polarized 1D PN junction and 0D PJ provide basic device components for future layertronic circuits based on electrical manipulation of the layer quantum number.

**Acknowledgements**

This work is supported by NSF CAREER Award 1944498. The material preparation and development of zero-twist angle double bilayer graphene was supported by NSF DMREF Award 1922165. The development of atomically-flat and strain-free gate-defined nanostructure Nanofabrication was conducted in the Minnesota Nano Center, which is supported by the National Science Foundation through the National Nano Coordinated Infrastructure Network, Award Number NNCI -1542202. Portions of the hexagonal boron nitride material used in this work were provided by K.W and T. T. Z.Z. is supported by a Stanford Science fellowship. K.W. and T.T. acknowledge support from the JSPS KAKENHI (Grant Numbers 20H00354, 21H05233 and 23H02052) and World Premier International Research Center Initiative (WPI), MEXT, Japan.

**References**

[1] Y. Cao, V. Fatemi, S. Fang, K. Watanabe, T. Taniguchi, E. Kaxiras, and P. Jarillo-Herrero, Unconventional superconductivity in magic-angle graphene superlattices, Nature **556**, 43 (2018).

[2] Y. Cao et al., Correlated insulator behaviour at half-filling in magic-angle graphene superlattices, Nature **556**, 80 (2018).

[3] M. Yankowitz, S. Chen, H. Polshyn, Y. Zhang, K. Watanabe, T. Taniguchi, D. Graf, A. F. Young, and C. R. Dean, Tuning superconductivity in twisted bilayer graphene, Science **363**, 1059 (2019).

[4] X. Lu et al., Superconductors, orbital magnets and correlated states in magic-angle bilayer graphene, Nature **574**, 653 (2019).

[5] A. L. Sharpe, E. J. Fox, A. W. Barnard, J. Finney, K. Watanabe, T. Taniguchi, M. A. Kastner, and D. Goldhaber-Gordon, Emergent ferromagnetism near three-quarters filling in twisted bilayer graphene, Science **365**, 605 (2019).

[6] M. Serlin, C. L. Tschirhart, H. Polshyn, Y. Zhang, J. Zhu, K. Watanabe, T. Taniguchi, L. Balents, and A. F. Young, Intrinsic quantized anomalous Hall effect in a moiré heterostructure, Science **367**, 900 (2020).

[7] Y. Saito, J. Ge, L. Rademaker, K. Watanabe, T. Taniguchi, D. A. Abanin, and A. F. Young, Hofstadter subband ferromagnetism and symmetry-broken Chern insulators in twisted bilayer graphene, Nat. Phys. **17**, 478 (2021).

[8] L. Wang et al., Correlated electronic phases in twisted bilayer transition metal dichalcogenides, Nat. Mater. **19**, 861 (2020).

[9] E. C. Regan et al., Mott and generalized Wigner crystal states in WSe2/WS2 moiré superlattices, Nature **579**, 359 (2020).

[10] X. Huang et al., Correlated insulating states at fractional fillings of the WS2/WSe2 moiré lattice, Nat. Phys. **17**, 6 (2021).

[11] J. Cai et al., Signatures of fractional quantum anomalous Hall states in twisted MoTe2, Nature **622**, 63 (2023).

[12] H. Park et al., Observation of fractionally quantized anomalous Hall effect, Nature **622**, 74 (2023).

[13] Y. Zeng, Z. Xia, K. Kang, J. Zhu, P. Knüppel, C. Vaswani, K. Watanabe, T. Taniguchi, K. F. Mak, and J. Shan, *Integer and Fractional Chern Insulators in Twisted Bilayer MoTe2*, http://arxiv.org/abs/2305.00973v3.

[14] G. W. Burg, J. Zhu, T. Taniguchi, K. Watanabe, A. H. MacDonald, and E. Tutuc, Correlated Insulating States in Twisted Double Bilayer Graphene, Phys. Rev. Lett. **123**, 197702 (2019).

[15] Y. Cao, D. Rodan-Legrain, O. Rubies-Bigorda, J. M. Park, K. Watanabe, T. Taniguchi, and P. Jarillo-Herrero, Tunable correlated states and spin-polarized phases in twisted bilayer–bilayer graphene, Nature **583**, 215 (2020).

[16] X. Liu et al., Tunable spin-polarized correlated states in twisted double bilayer graphene, Nature **583**, 7815 (2020).

[17] R. Su, M. Kuiri, K. Watanabe, T. Taniguchi, and J. Folk, Superconductivity in twisted double bilayer graphene stabilized by WSe2, Nat. Mater. **22**, 11 (2023).

[18] M. He, J. Cai, Y.-H. Zhang, Y. Liu, Y. Li, T. Taniguchi, K. Watanabe, D. H. Cobden, M. Yankowitz, and X. Xu, *Symmetry-Broken Chern Insulators in Twisted Double Bilayer Graphene*, http://arxiv.org/abs/2109.08255v2.

[19] P. Rickhaus et al., Correlated electron-hole state in twisted double-bilayer graphene, Science **373**, 1257 (2021).

[20] M. Yamamoto, Revealing Dielectric Screening in Twisted Graphene Devices, JPSJ News Comments **19**, 02 (2022).

[21] H. Yoo et al., Atomic and electronic reconstruction at the van der Waals interface in twisted bilayer graphene, Nat. Mater. **18**, 5 (2019).

[22] T. Latychevskaia, C. Escher, and H.-W. Fink, Moiré structures in twisted bilayer graphene studied by transmission electron microscopy, Ultramicroscopy **197**, 46 (2019).

[23] T. A. de Jong, T. Benschop, X. Chen, E. E. Krasovskii, M. J. A. de Dood, R. M. Tromp, M. P. Allan, and S. J. van der Molen, Imaging moiré deformation and dynamics in twisted bilayer graphene, Nat Commun **13**, 70 (2022).

[24] X. Zhang et al., Gate-tunable Veselago interference in a bipolar graphene microcavity, Nat Commun **13**, 6711 (2022).

[25] W. Ren, X. Zhang, Z. Zhu, M. Khan, K. Watanabe, T. Taniguchi, E. Kaxiras, M. Luskin, and K. Wang, Electron Collimation in Twisted Bilayer Graphene via Gate-Defined Moiré Barriers, Nano Lett. **24**, 12508 (2024).

[26] E. McCann, D. S. L. Abergel, and V. I. Fal'ko, The low energy electronic band structure of bilayer graphene, Eur. Phys. J. Spec. Top. **148**, 91 (2007).

[27] L. Tarruell, D. Greif, T. Uehlinger, G. Jotzu, and T. Esslinger, Creating, moving and merging Dirac points with a Fermi gas in a tunable honeycomb lattice, Nature **483**, 7389 (2012).

[28] K. K. Gomes, W. Mar, W. Ko, F. Guinea, and H. C. Manoharan, Designer Dirac fermions and topological phases in molecular graphene, Nature **483**, 7389 (2012).

[29] J. Simon and M. Greiner, A duo of graphene mimics, Nature **483**, 7389 (2012).

[30] T. Uehlinger, G. Jotzu, M. Messer, D. Greif, W. Hofstetter, U. Bissbort, and T. Esslinger, Artificial Graphene with Tunable Interactions, Phys. Rev. Lett. **111**, 185307 (2013).

[31] K. Davydov, X. Zhang, W. Ren, M. Coles, L. Kline, B. Zucker, K. Watanabe, T. Taniguchi, and K. Wang, Easy-to-configure zero-dimensional valley-chiral modes in a graphene point junction, Science Advances **10**, eadp6296 (2024).

[32] K. Wang, M. M. Elahi, L. Wang, K. M. M. Habib, T. Taniguchi, K. Watanabe, J. Hone, A. W. Ghosh, G.-H. Lee, and P. Kim, Graphene transistor based on tunable Dirac fermion optics, Proceedings of the National Academy of Sciences **116**, 6575 (2019).

[33] D. Rodan-Legrain, Y. Cao, J. M. Park, S. C. de la Barrera, M. T. Randeria, K. Watanabe, T. Taniguchi, and P. Jarillo-Herrero, Highly tunable junctions and non-local Josephson effect in magic-angle graphene tunnelling devices, Nat. Nanotechnol. **16**, 769 (2021).

[34] R. V. Gorbachev et al., Strong Coulomb drag and broken symmetry in double-layer graphene, Nature Phys **8**, 896 (2012).

[35] X. Liu, K. Watanabe, T. Taniguchi, B. I. Halperin, and P. Kim, Quantum Hall drag of exciton condensate in graphene, Nature Phys **13**, 746 (2017).

[36] F. Escudero, F. Arreyes, and J. S. Ardenghi, Coulomb drag between two graphene layers at different temperatures, Phys. Rev. B **106**, 245414 (2022).

[37] K. Wang, A. Harzheim, T. Taniguchi, K. Watanabei, J. U. Lee, and P. Kim, Tunneling Spectroscopy of Quantum Hall States in Bilayer Graphene p- n Junctions, Physical Review Letters **122**, 146801 (2019).

[38] Y. Saito, J. Ge, K. Watanabe, T. Taniguchi, and A. F. Young, Independent superconductors and correlated insulators in twisted bilayer graphene, Nat. Phys. **16**, 926 (2020).

[39] A. K. Geim, Graphene: Status and Prospects, Science **324**, 1530 (2009).

[40] L. Wang et al., One-Dimensional Electrical Contact to a Two-Dimensional Material, Science **342**, 614 (2013).

[41] C. R. Dean et al., Boron nitride substrates for high-quality graphene electronics, Nature Nanotech **5**, 10 (2010).

[42] L.-J. Yin, K.-K. Bai, W.-X. Wang, S.-Y. Li, Y. Zhang, and L. He, Landau quantization of Dirac fermions in graphene and its multilayers, Front. Phys. **12**, 127208 (2017).

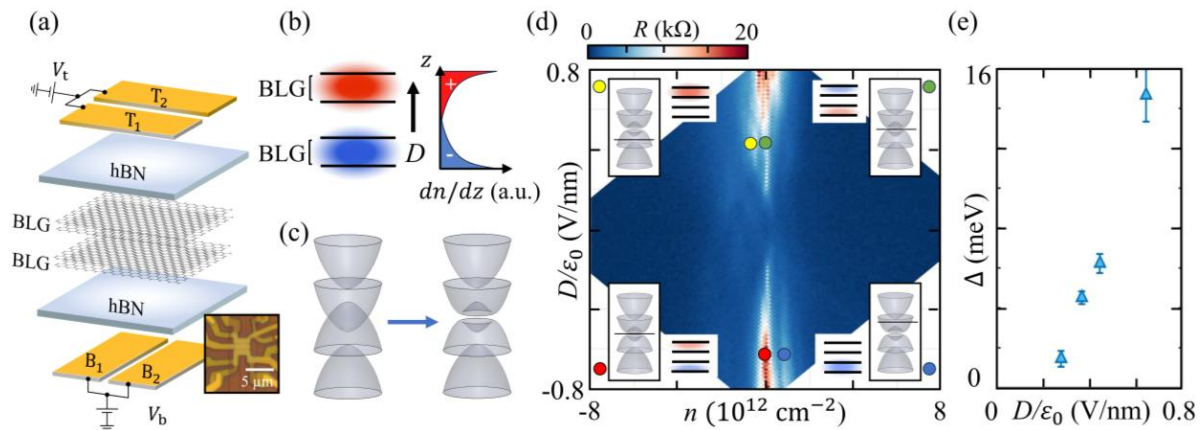

**Figure 1. Layer-polarization in B0B.** (a) Device schematics: hBN-encapsulated double-BLG with zero twist and a pair of top and bottom gates. A uniform 2DEG and perpendicular D field can be applied by applying same voltages to adjacent gates. Inset: Optical image of a typical device. (b) This separates electrons (blue) and holes (red), and results in layer polarized charge distribution (right), and (c) crossing between BLG bands due to layer-polarized band offset. (d) four-probe longitudinal resistance versus total carrier density and $D$ fields. Resistance peaks correspond to layer-polarized (yellow, blue) and charge-neutral states (red and green). (e) Extracted band gap of B0B as a function of $D$.

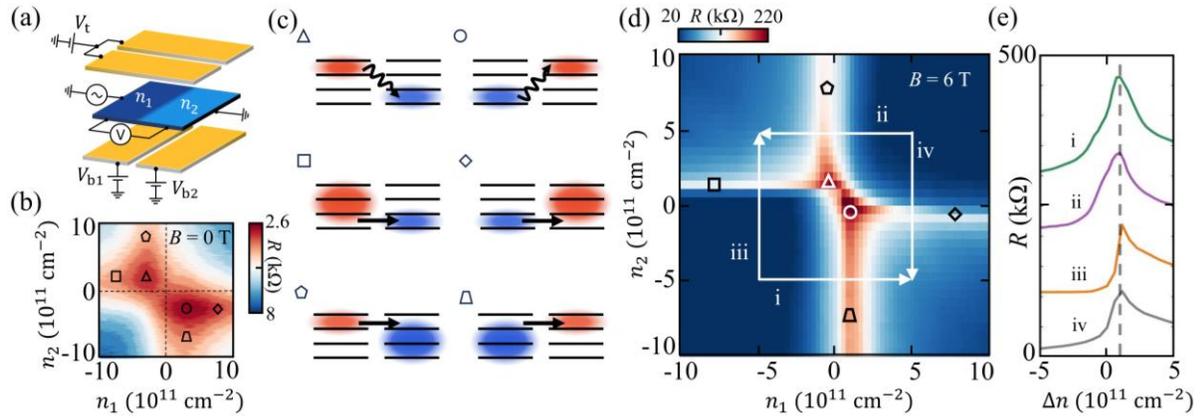

**Figure 2. Layer-polarized transport in 1D PN-junction.** (a) Schematic of the 1D junction configuration. (b) Four-probe resistance measured while the bottom gates are separately biased to independently tune the carrier densities $n_1$ and $n_2$ of two regions above at $V_t$= -7.8V and B = 0T. The resistance peaks exhibits a broken-cross shape, due to (c) tunneling (curved arrow) or ballistic (arrow) transport across layer-polarized PN junction, (d) which can be enhanced by finite magnetic field B = 6T. (e) 1D cuts along white arrows in (d), exhibiting the same shift from charge neutrality, consistent with layer-polarization in B0B.

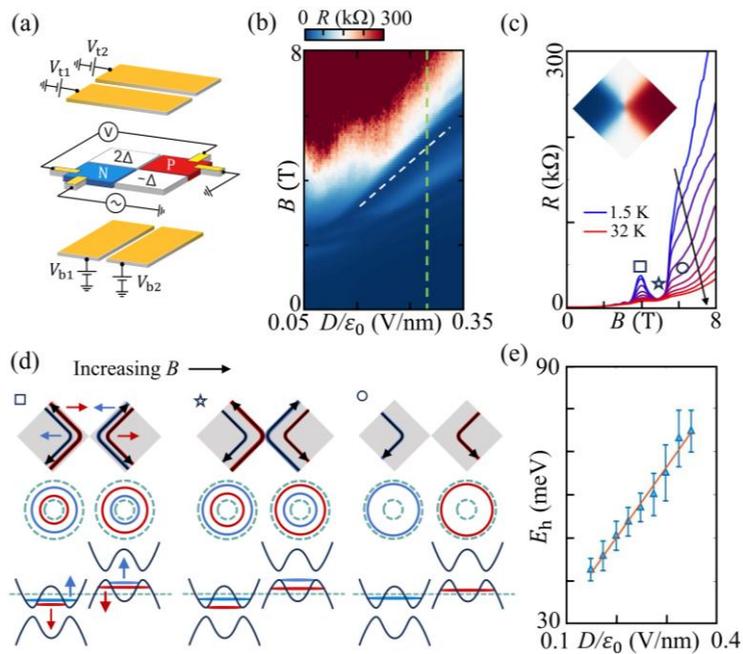

**Figure 3. Layer-polarized Magneto Transport via a 0D Point Junction.** (a) Schematic of a 0D PJ device configuration. four independent local gate voltages electrostatically separate the device into four regions: doped regions with opposite carrier types, and insulating regions with opposite D field. (b) Measured four-probe resistance across the PJ as function of B and D fields. The white dash line marks the position of the last resistance dip. Along $D/\varepsilon_0$=0.3 V/nm (green dashed), (c) resistance versus $B$ at different temperature. (d) Top: Spatial distribution of N-type (blue) and P-type (red) QH states near the PJ; arrows indicate the movement direction with increased B. Middle and Bottom: The P-type (red) and N-type (blue) Landau level with respect to two Fermi surfaces (dashed) and band structure (solid black). (e) The estimated size of layer-polarized band crossing as a function of D field.

Supplementary Materials

# Layer-polarized Transport via Gate-defined 1D and 0D PN Junctions in Double Bilayer Graphene

Wei Ren[1], Xi Zhang[1], Shiyu Guo[1], Jeongsoo Park[1], Jack Tavakley[1], Daochen Long[1], Kenji Watanabe[2], Takashi Taniguchi[3], Ke Wang[1*]

[1]School of Physics and Astronomy, University of Minnesota, Minneapolis, Minnesota 55455, USA
[2]Research Center for Electronic and Optical Materials, National Institute for Materials Science, 1-1 Namiki, Tsukuba 305-0044, Japan
[3]Research Center for Materials Nanoarchitectonics, National Institute for Materials Science, 1-1 Namiki, Tsukuba 305-0044, Japan

**S1. Sample Preparation and Device Fabrication**

A pair of parallel-aligned metal gates ($B_1$ and $B_2$, with a ~100 nm separation between them) consisting of Cr/Pd-Au alloy (1 nm / 7 nm) are deposited on a $SiO_2$ (285 nm) / Si (doped) substrate. The Pd-Au alloy (40% Pd / 60% Au) is chosen to reduce the surface roughness when it is compared to conventional pure Au deposition. The gates are subsequently annealed in a high-vacuum environment at 350℃ for 10 minutes to remove surface residue.

The zero-twist angle double bilayer graphene (B0B) stacks are made by the 'cut and tear' method [1]. Hexagonal boron nitride (hBN, 40 ~ 80 nm) and bilayer graphene (BLG) flakes are exfoliated [2] and characterized by atomic force microscope (AFM) to be atomically clean. A single piece of BLG is cut into three individual pieces with the same lattice orientation by AFM. With the help of a poly polypropylene carbonate (PPC) and polydimethylsiloxane (PDMS) stamp on a glass slide [3], we pick up the first piece of hBN and the two pieces of the precut graphene consecutively, aiming at zero twist angle. The second piece of hBN is picked up to encapsulate the stack, and the whole stack is transferred onto the prepatterned bottom gates at 100℃ [4]. Then the sample is rinsed in acetone and isopropanol to remove the PPC residue on the top hBN surface.

For sample S1 [inset in Fig. 1(a)], three parallel-aligned metal gates ($T_1$, $T_2$, and $T_3$, with a ~100 nm separation between them) consisting of Cr/Pd/Au (1 nm / 5 nm / 14 nm) are deposited on the top surface of the hBN encapsulated B0B sample. The orientation of $T_1$, $T_2$, and $T_3$ is perpendicular to the three redeposited bottom metal gates $B_1$, $B_2$, and $B_3$. For sample S2, a global top gate consisting of

Cr/Pd/Au (1 nm / 5 nm / 14 nm) is deposited on the top surface of the sample instead of a pair of parallel-aligned metal gates ($T_1$ and $T_2$). Electrical contacts to gates and ohmic contacts to 1D boundaries of the B0B sample are made by electron-beam lithography, dry-etching and subsequent metal deposition (Cr/Pd/Au, 1 nm/5 nm/ 180 nm). A final round of dry-etching with an electron-beam lithography etch mask defines the lateral geometry of the devices.

Since the thickness of hBN is comparable to the separation between the two local bottom (top) gates, the electrostatic potential profile at the graphene surface (40 ~ 80 nm away from the gates) is expected to be uniform when the same voltage is applied to both bottom (top) gates, despite the gate separation. Therefore, two (or three) bottom (top) gates can function as a global bottom (top) gate when needed, allowing the B0B devices to be configured into either a dual-global-gated device, a PN-junction device, or a 0D PJ device. As an example, sample S1 can in principle be configured into a 3×3 grid where the carrier density and gap size in each region are determined by the top and bottom gate voltage configuration. For 1D PN junction measurements, two of the three bottom metal gates (such as $B_2$ and $B_3$) can be connected to the same DC voltage source to serve as a single bottom gate, while all three top gates are held at the same DC voltage to function as a global top gate. Alternatively, the configuration can be reversed with two top gates (such as $T_2$ and $T_3$) connected as a single top gate while all three bottom gates act as a global bottom gate. In the 0D PJ measurements, the device is configured into multiple 2×2 grids by setting pairs of adjacent top or bottom gates to identical voltages. While the sample design allows for four possible PJ devices in principle, one non-functional 1D contact (to the B0B sample) at the bottom right corner of sample S1 limits the measurements to three operational PJ devices (PJ#1, PJ#2, and PJ#3), which are characterized using four-terminal measurements.

Data measured at the global dual-gated and PN-junction configurations are collected by a standard SR830 lock-in amplifier with an alternating current excitation of 10 nA at 17.777778 Hz applied through the device. Data measured at the 0D PJ configuration are collected with an alternating voltage excitation of 1 mV at 17.777778 Hz. A Digital-to-Analog Converter (DAC) and/or Yokogawa DC voltage sources are used to apply voltages on the local metal gates. Sample S1 was characterized by an Oxford TeslatronPT cryostat at a base temperature of 1.5 K. Sample S2 measurements were performed under different conditions: (1) at zero magnetic field using a Montana cryostat at a base temperature of 4 K, and (2) under applied magnetic fields using a Physical Property Measurement System (PPMS) at 20 K.

## S2. Characterization of Band Gap via Temperature Dependence

The temperature dependence of four-probe longitudinal resistance can be used to determine the size of the band gap due to the inter-layer tunneling between the two BLG sheet. The measurement is conducted in sample S1 in the dual-global-gated configuration at zero doping and different D fields. The following equation is used to fit the measured longitudinal resistance as a function of temperature $T$:

$$R \propto e^{\frac{-\Delta}{2k_B T}}. \tag{1}$$

The gap monotonically increases with the displacement fields up to ~ 15 meV at $D/\varepsilon_0$ = -0.64 V/nm.

## S3. Coulomb-drag Measurements via 1D PN-junction

Coulomb-drag measurements are taken to confirm the layer polarization, it can also serve as an ideal platform to study the Coulomb-drag and the exciton condensation in the strongest coupling regime with a separation of only a few atoms compared to the traditional hBN spacer device [5–7]. The non-local measurement setup is shown in Fig. S1(a), where the drag signal is detected in a different gate region ($B_2$) than the driving signal ($B_1$). Fig. S1(b) shows the drag signal in Sample S1 as functions of $n_1$ and $n_2$ at $V_t$ = +4 V and $B$ = –4 T. In addition to the expected peak at zero doping (or along $n_1 = n_2$) that is attributable to dissipative transport when the sample is uniform and charge neutral, an off-diagonal (along $n_1 \approx -n_2$) peak in the drag signal is observed. This feature corresponds to the case where the P-type carrier ($n_1 < 0$) is polarized in the top BLG of region 1 (on top of $B_1$), while the N-type carrier ($n_2 > 0$) is polarized in the bottom BLG of region 2 (on top of $B_2$). The observed Coulomb-drag signal confirms the formation of 1D PN junction both laterally and vertically defined.

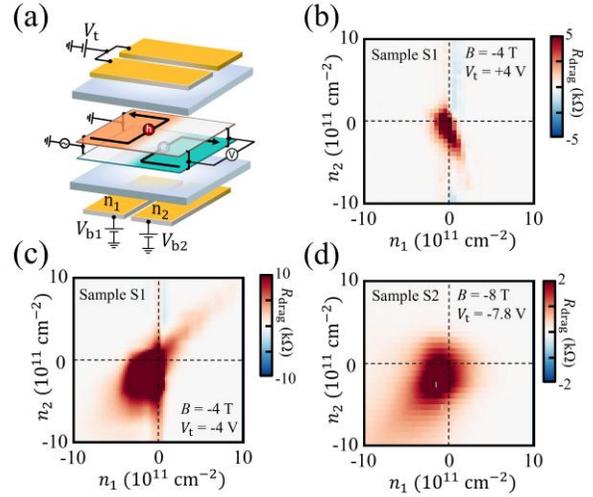

**Figure S1. Coulomb Drag Measurements via a 1D PN-junction.** (a) The schematic image of the drag measurement configuration where V denotes the drag signal. Due to the symmetry-protected interlayer tunneling and the strong electrostatic screening, the B0B device provides an unprecedented platform to study the Coulomb drag. (b) shows the non-local drag signals at B = -4 T. The strong Coulomb drag response appears near the interlayer charge imbalance configurations [Fig. 2(c), triangle and circle].

However, the drag signal cannot always be clearly resolved as it is near the blob of high background resistance at the charge neutrality point. Figs. S1(c), (d) are the measured non-local signal in another gate-configuration in sample S1 ($V_t$ = -4 V and $B$ = –4 T) and sample S2 ($V_t$ = -7.8 V and $B$ = –8 T). Although signature of non-local signal is observed in the $n_1 \cdot n_2 < 0$ regime, which indicate the presence of Coulomb drag, it cannot be clearly distinguished with the background signal. Future device designs and measurement schemes are needed to verify the presence of Coulomb-drag.

## S4. Controlled Configuration to Verify Resistance Dip Mechanisms

The existence of the resistance dip relies on the direct contact between the inner QH edge states. To further verify this physics picture, the device is configured into $(-V_0, -2V_0, +2V_0, +V_0)$ as depicted in Fig. S2(a), where the bottom left (top right) region remains N (P) doped, but the displacement fields applied to the other two gapped regions are now of the same direction instead of opposite. In contrast to the band gap closing at the PJ in the previous configuration, the band gap does not close at the controlled PJ configuration; instead, a finite tunneling barrier is established between the P and N regions [Fig. S2(c), inset]. In the absence of direct contact between inner-band QH edge states, local resistance minima are ill-defined and exhibit strong T-dependence at magnetic fields beyond $B$ = 4 T. This is expected, as conductance is limited by the tunneling rate between the innermost QH edge states, which is strongly enhanced by elevating the temperature. The data in the controlled PJ configuration further confirms our physics picture.

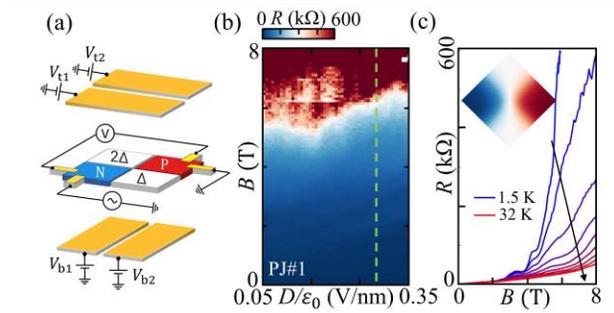

**Figure S2. Magneto-transport via a Controlled 0D Point Junction Configuration.** (a) The schematic image of the $(-V_0, -2V_0, +2V_0, +V_0)$ configuration. (b) Corresponding four-probe resistance across the PJ versus $B$ and $D$ fields. (c) 1D cut along $D$ = 0.3 V/nm (green dashed line in (b)) at different temperatures. The resistance increases monotonically with $B$. (Inset) Schematic of electrostatic profile in the PJ device, where a gap is present at the center of the PJ.

## S5. Additional Data on 1D PN Junction Characterization

Here we provide additional data on characterizing the electrostatic screening effect via the 1D PN junction to supplement the main data. Fig. 2(b) is the zero magnetic field result from sample S2, and Fig. 2d is from sample S1 at $B$ = 6T.

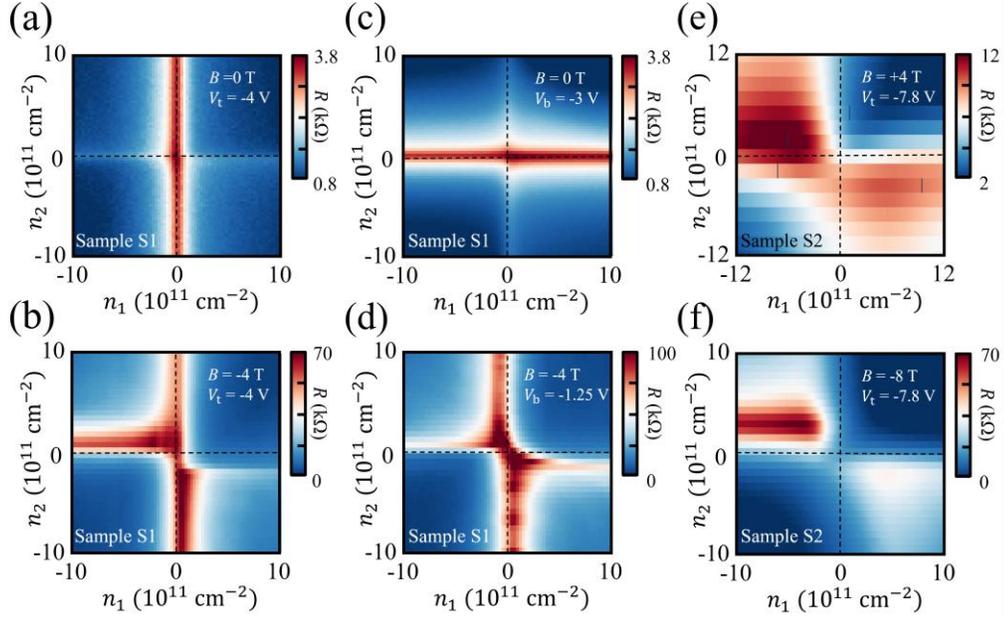

**Figure S3. Characterization on Electrostatic Screening Effect via 1D PN Junction in Sample S1 and S2.** (a-b) Four-probe longitudinal resistance across the 1D PN junction in sample S1 as functions of $n_1$ and $n_2$ at (a) $B = 0$ T, $V_T = -4$ V (b) $B = -4$ T, $V_T = -4$ V. (c-d) Adopting the global bottom gate configuration, four-probe longitudinal resistance across the 1D PN junction in sample S1 as functions of $n_1$ and $n_2$ at (d) $B = 0$ T, $V_B = -3$ V (e) $B = -4$ T, $V_B = -1.25$ V. (e-f) Four-probe longitudinal resistance across the 1D PN junction in sample S2 as functions of $n_1$ and $n_2$ at $B = 4$ T, $V_T = -4$ V.

As sample S1 has multiple top and bottom gates, there are more possible configurations available for characterizing the electrostatic screening via a 1D PN junction. The first configuration [Figs. S3(a), (b)] is to have all top gates parked at the same voltages $V_t$, while the two effective bottom gate (see section S1) voltages $V_{B1}$ and $V_{B2}$ vary independently. The source, drain contacts and voltage probes are set up correspondingly into a similar configuration as in Fig. 2(a). The second configuration [Figs. S3(c), (d)) is to swap the role of top and bottom gates, i.e., park all bottom gates at the same voltage $V_b$, while vary the two effective top gate voltages $V_{t1}$ and $V_{t2}$ independently. In either configuration, the broken-cross signature at 0 T [Figs. S3(a), (c)] is not as well resolved as in sample S2 [Fig. 2(b)]. It may be because the intrinsic inter-layer tunneling is stronger in sample S1, which suppresses the zero magnetic field electrostatic screening effect. When a finite $B$ is applied, discrete Landau levels (LL) in each BLG layer develop. The charge carriers in opposite layers decouple from each other unless the LL in the top and bottom BLG are in resonance. The inter-layer tunneling between the two BLG layers is, therefore, suppressed, and the electrostatic screening is enhanced. The presence of the unconventional broken-cross shaped resistance peaks [Figs. S3(b), (d)] suggest the presence of strong charge imbalance in sample S1 with the help of a $B$ field.

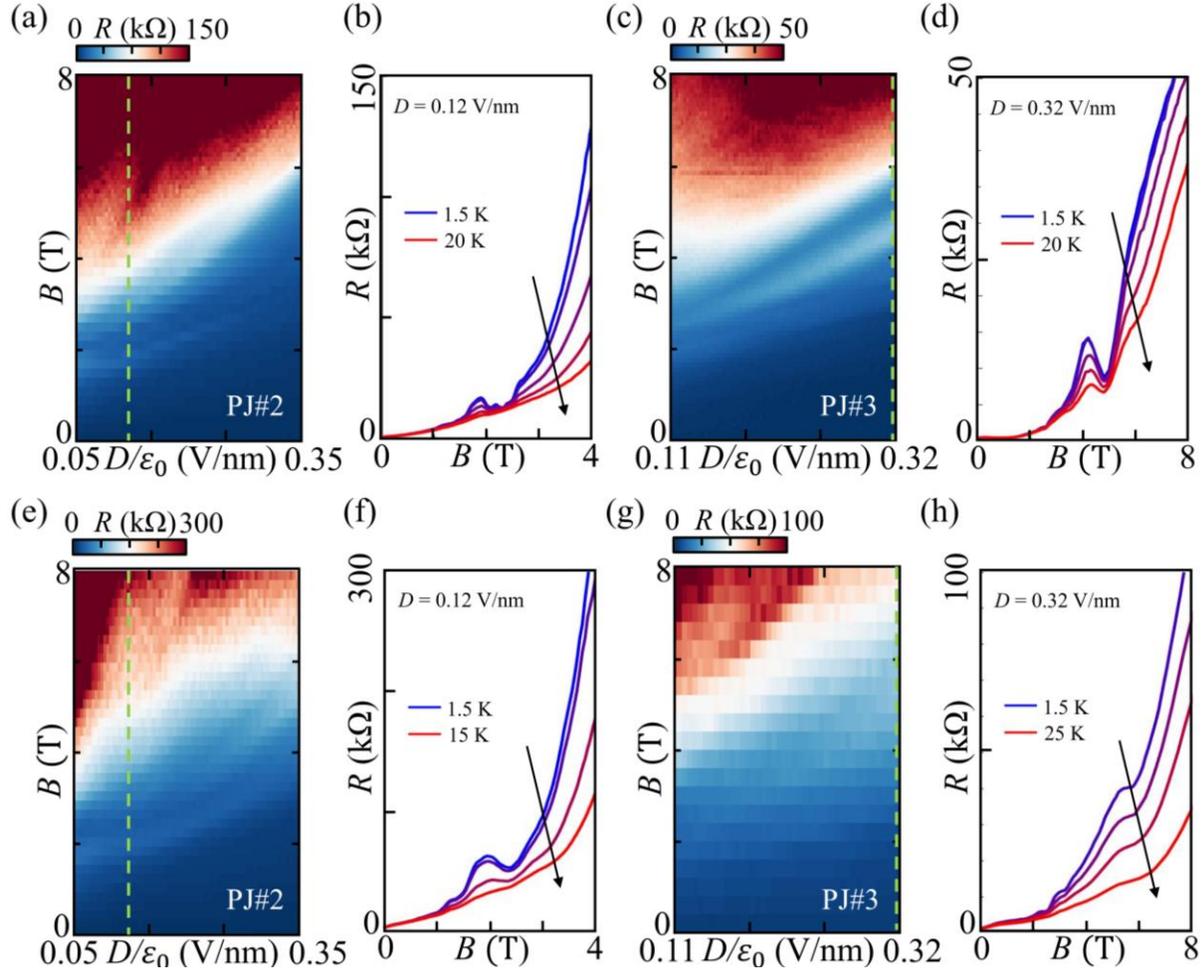

**Figure S4. 0D PJ Characterization on PJ#2 and PJ#3.** (a-b) 0D PJ characterization on PJ#2 in the (+$V_0$, -2$V_0$, +2$V_0$, -$V_0$) configuration. (a) Four-probe resistance across the PJ as functions of $B$ and $D$ fields. Along $D/\varepsilon_0$=0.12 V/nm (green dashed line), (b) resistance versus $B$ at difference temperatures with a resistance dip with minimal temperature dependence at $B$ = 2.3 T. (c-d) 0D PJ characterization on PJ#3 in the (+$V_0$, -2$V_0$, +2$V_0$, -$V_0$) configuration. (e-f) 0D PJ characterization on PJ#2 in the (-$V_0$, -2$V_0$, +2$V_0$, +$V_0$) configuration. (e) Four-probe resistance across the PJ as functions of $B$ and $D$ fields. Along $D/\varepsilon_0$=0.12 V/nm (green dashed line), (f) resistance versus $B$ at difference temperatures with stronger the temperature dependence than in the other gate-configuration. (g-h) 0D PJ characterization on PJ#3 in the (+$V_0$, -2$V_0$, +2$V_0$, -$V_0$) configuration.

Fig. S3 (e)-(f) are the four-probe resistance measured via the 1D PN junction configuration at $B$ = +4 T and $B$ = -8 T, respectively. The top gate voltage $V_t$ = -7.8 V in both cases. Due to the formation of LLs, the lightly-doped BLG layers become more insulated. The resistance peaks are significantly enhanced at the inter-layer charge imbalanced configuration [Fig. 2(c), triangle and circle].

**S6. 0D PJ Characterization in Additional Devices**

The data in Figs. 3 and S2 are taken from PJ#1 in sample S1. Two additional point junctions (PJ#2 and PJ#3) in sample S1 are measured in both ($+V_0$, $-2V_0$, $+2V_0$, $-V_0$) and ($-V_0$, $-2V_0$, $+2V_0$, $+V_0$) configurations. In both point junctions, the ($-V_0$, $-2V_0$, $+2V_0$, $+V_0$) configurations [Figs. S4(e)-(g)] have an overall higher resistance compared to the other configuration [Figs. S4(a)-(d)]. This is consistent with the expectation that the ($-V_0$, $-2V_0$, $+2V_0$, $+V_0$) configuration features a tunnel barrier with finite width, significantly contributing to the measured resistance across the PJ. In contrast, the width of the tunnel barrier in the other configuration gradually drops to zero as it is closer to the center of the PJ, facilitating the tunneling of QH edge state between the N and P regions.

In PJ#2, the ($-V_0$, $-2V_0$, $+2V_0$, $+V_0$) configurations show similar $B$ field, $D$ field and temperature dependence (figures S4a-b) as PJ#1 discussed in the main text. At a constant $D$ field, a resistance dip with minimal temperature dependence is observed. When $D/\varepsilon_0 = 0.12$ V/nm, the resistance dip is found around $B = 2.3$ T. The corresponding $B$ value of the resistance dip also increases with $D$ field as expected. In the ($-V_0$, $-2V_0$, $+2V_0$, $+V_0$), although the monotonic behavior in $B$ field is not observed as in PJ#2, the temperature dependence of the measured resistance at a given $D$ field is still stronger than the previous configurations in the same PJ device. The non-monotonic behavior may be because the tunnel barrier in PJ#2 is weaker or narrower compared to PJ#1, the quantum Hall edge states on the two sides are partially hybridized even with the presence of a tunnel barrier. In PJ#3, lower overall resistance is measured compared to PJ#1 in either configure, but the essential features are qualitatively reproduced.

**S7. Characterization of Energy Scale of Sombrero-like Band**

Here we provide a zeroth-order estimation on the energy scale of the "Sombrero" by assuming that the band structure of two BLG stay relatively unaltered due to the presence of another BLG, except for the states near the crossing point of two BLG bands [Fig. 1(c)]. This includes that the property of the Landau levels originating from the two BLG remain approximated unaltered. A more accurate description involves a detailed calculation of the band structure, which is out of the scope of this work.

The energy of Landau levels (LLs) in biased BLG follows[5]:

$$E_0 = E_C + \frac{\zeta U}{2},$$

$$E_1 = E_C + \zeta U/2(1 - z),$$

$$E_n = E_C \pm \sqrt{(\hbar \omega_c)^2[n(n-1)] + (U/2)^2} - \zeta z U/4, \qquad n = 2, 3, 4 \ldots$$

Where $E_C$ is the energy of the charge-neutrality point, $U$ is the interlayer bias, $\zeta = +/-$ are valley indices, and $\omega_c = \frac{eB}{m^*}$ is the cyclotron frequency, where $e$ is the electron charge, and $m^*$ is the effective electron mass in BLG. For $B < 8$ T, $z = \frac{2\hbar\omega_c}{t_\perp} \ll 1$, which is negligible.

From the above expression, $n = 0, 1$ LLs are relatively unchanged as the $B$ field increases and staying at the bottom of the Sombrero-like band. As $B$ increases, the $n = 2$ LL shifts upward in energy, eventually merging with the top of the "Sombrero" at high fields [Fig. 3(d)], till it disappears. Thus, the energy scale of the "Sombrero" is estimated as the energy difference between the lower branch of $n = 2$ LL and $n = 0$ (or $n = 1$) Landau levels at the $B^*$, where $B^*$ is the corresponding magnetic field of the last resistance dip measured in Fig. 3(b) (while-dash line):

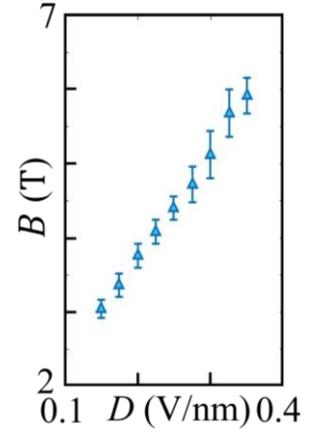

**Figure S5. Characterizing Sombrero-like Band Dispersion.** $B^*$ value of the last oscillation dip measured at different displacement fields, extracted from Fig. 3(b).

$$E_h = E_2 - E_0 = \sqrt{\left(\frac{eB}{m^*}\right)^2 + \left(\frac{U}{2}\right)^2} - \frac{U}{2} \qquad (2)$$

Fig. S5 is the $B^*$ value of the last oscillation dip measured at different displacement field, extracted from Fig. 3(b). Take $m^* = 0.04 m_e$, the height of the "Sombrero" as a function of the displacement field is calculated using the above equation (2) and plotted in Fig. 3(e).

## References

[1] Y. Saito, J. Ge, K. Watanabe, T. Taniguchi, and A. F. Young, Independent superconductors and correlated insulators in twisted bilayer graphene, Nat. Phys. **16**, 926 (2020).

[2] A. K. Geim, Graphene: Status and Prospects, Science **324**, 1530 (2009).

[3] L. Wang et al., One-Dimensional Electrical Contact to a Two-Dimensional Material, Science **342**, 614 (2013).

[4] C. R. Dean et al., Boron nitride substrates for high-quality graphene electronics, Nature Nanotech **5**, 10 (2010).

[5] R. V. Gorbachev et al., Strong Coulomb drag and broken symmetry in double-layer graphene, Nature Phys **8**, 896 (2012).

[6] X. Liu, K. Watanabe, T. Taniguchi, B. I. Halperin, and P. Kim, Quantum Hall drag of exciton condensate in graphene, Nature Phys **13**, 746 (2017).

[7] F. Escudero, F. Arreyes, and J. S. Ardenghi, Coulomb drag between two graphene layers at different temperatures, Phys. Rev. B **106**, 245414 (2022).

[8] L.-J. Yin, K.-K. Bai, W.-X. Wang, S.-Y. Li, Y. Zhang, and L. He, Landau quantization of Dirac fermions in graphene and its multilayers, Front. Phys. **12**, 127208 (2017).